\documentclass[%
 aip,
 amsmath,amssymb,
 reprint,%
]{revtex4-1}

\usepackage{graphicx}
\usepackage{dcolumn}
\usepackage{bm}
\usepackage{textcomp}
\usepackage{gensymb}
\usepackage{tabularx}
\usepackage{color}
\usepackage{float}
\usepackage{bm}
\usepackage{amsmath}
\usepackage{amssymb}
\usepackage{verbatim}

\newcolumntype{Y}{>{\centering\arraybackslash}X}

\begin{document}
\title{Experimental and Theoretical Investigation of Resonances in Low-Energy NO-H$_2$ collisions} 

\author{Quan Shuai}
\email[]{Who contributed equally to this work.}
\author{Tim de Jongh}
\email[]{Who contributed equally to this work.}
\author{Matthieu Besemer}
\author{Ad van der Avoird}
\author{Gerrit C. Groenenboom}
\author{Sebastiaan Y.T. van de Meerakker}
\email[]{Corresponding author: basvdm@science.ru.nl}
\affiliation{Radboud University, Institute for Molecules and Materials, Heijendaalseweg 135, 6525 AJ Nijmegen, the Netherlands}


\date{\today}

\begin{abstract}
The experimental characterization of scattering resonances in low energy collisions has proven to be a stringent test for quantum chemistry calculations. Previous measurements on the NO-H$_2$ system at energies down to $10$ cm$^{-1}$ challenged the most sophisticated calculations of potential energy surfaces available. In this report, we continue these investigations by measuring the scattering behavior of the NO-H$_2$ system in the previously unexplored $0.4 - 10$ cm$^{-1}$ region for the parity changing de-excitation channel of NO. We study state-specific inelastic collisions with both \textit{para}- and \textit{ortho}-H$_2$ in a crossed molecular beam experiment involving Stark deceleration and velocity map imaging. We are able to resolve resonance features in the measured integral and differential cross sections. Results are compared to predictions from two previously available potential energy surfaces and we are able to clearly discriminate between the two potentials. We furthermore identify the partial wave contributions to these resonances, and investigate the nature of the differences between collisions with \textit{para}- and \textit{ortho}-H$_2$. Additionally, we tune the energy spreads in the experiment to our advantage to probe scattering behavior at energies beyond our mean experimental limit.
\end{abstract}

\pacs{}

\maketitle 

\section{Introduction}

Scattering resonances are among the most subtle phenomena observed in molecular collisions and therefore serve as an extremely sensitive probe for theoretical descriptions of molecular interactions.\cite{Levine:reaction-dynamics,Naulin:IRPC33:427} They appear only at low collision energies, i.e., conditions under which the de Broglie wavelength of the particles is of the same order of magnitude as the characteristic distance of the interaction. Each resonance corresponds to the formation of a transiently bound Van der Waals complex of the interacting particles, and its characteristics are influenced by minute details in the underlying potential energy surface (PES). A resonance manifests itself most notably as a narrow peak in the energy dependence of an integral cross section (ICS) where the probability of scattering increases when the collision energy approaches that of a quasi-bound state. It can further be characterized by analyzing the corresponding rapid variations in the differential cross section (DCS), revealing the underlying wavelike nature of the collision event.

The experimental observation of scattering resonances demands access to low collision energies as well as a high energy resolution, a challenge in which the molecular beam technique has been an essential tool. The first measurements of resonances were performed in a crossed molecular beam apparatus and involved elastic collisions between hydrogen and mercury atoms.\cite{Schutte:PRL29:979} Later on, similar techniques were used to study resonances in collisions of hydrogen atoms and molecules with a variety of scattering partners.\cite{Grover:ChemPhysLett:48:24,Toennies:JCP71:614} These observations were in part made possible by the advantageous kinematics for the respective systems,\cite{Chandler:JCP132:110901} and it would take years until experimental progress allowed for the observation of resonances in different systems.

The past two decades have seen a resurgence in experimental studies on resonances. Several crossed molecular beam experiments have been performed which aimed to observe signatures of scattering resonances in the seminal F + H$_2$ reaction.\cite{Skodje:PRL85:1206,Shiu:PRL92:103201,Qiu:Science311:1440,Dong:Science327:1501,Wang:Science342:1499} Furthermore, resonances in Penning ionization reactions have been studied with the merged beam technique. In these experiments, curved magnetic and electrostatic guides were used to attain zero degree scattering angles and reach energies down to a few milikelvin.\cite{Henson2012,Lavert-Ofir:NatChem6:332,Jankunas2015,Shagam2015,Klein:NatPhys13:35} The first observation of scattering resonances in the ICS of state-to-state rotationally inelastic collisions was made in a crossed molecular beam apparatus in which cryogenically cooled beams collided at small scattering angles. This allowed for thorough verification of theoretical models for a host of astrochemically relevant systems, involving species such as O$_2$, CO, and H$_2$.\cite{Chefdeville:Science341:06092013,Bergeat:NatChem7:349,Chefdeville:AJL799:L9}

In recent years, full characterization of resonances in inelastic scattering has become possible by combining the Stark deceleration and Velocity Map Imaging (VMI) techniques in a crossed beam experiment. Stark deceleration provides a robust method for the preparation of velocity selected, state pure packets of molecules with narrow spatial and velocity distributions. It thereby provides an adjustable collision energy with small energy spreads. The VMI detection allows for state-selective detection of the collision products, and enables measurements of both the ICS and the DCS. This approach led to full identification and characterization of resonance structures in NO-He and NO-H$_2$ collisions.\cite{Vogels:SCIENCE350:787,Vogels:NatChem10:435,Jongh2020}

For the NO-He and NO-H$_2$ systems, the spatially degenerate nature of the electronic states of the NO radical dramatically increases the complexity of quantum chemical calculations describing these experiments. Open-shell radical systems therefore serve as a crucial test for such computational methods, and the recent measurements of resonances in these systems have challenged the validity of even the most sophisticated quantum chemistry calculations.
Experimental observation of resonances in the NO-H$_2$ system at energies down to 10 cm$^{-1}$ enabled the discernment between two potentials, both constructed with the coupled-cluster method with the inclusion of single and double
excitations and perturbative treatment of triples -- better known as the CCSD(T) method.\cite{Vogels:NatChem10:435}
For NO-He, recent characterization of resonances at energies down to 0.2 cm$^{-1}$ required even higher levels of theory. Only a CCSDT(Q) potential -- in which full treatment of triple and perturbative treatment of quadruple excitations was included -- was able to capture the experimentally observed resonance features in the 0.2 cm$^{-1}$ - 8.5 cm$^{-1}$ range.\cite{Jongh2020}

Here, we report new measurements of resonances in state-to-state inelastic NO-H$_2$ collisions, extending the experimentally probed energy range to collision energies as low as 0.4 cm$^{-1}$. We measured DCSs and ICSs in a crossed molecular beam experiment involving Stark decelerated NO molecules and cryogenically cooled hydrogen molecules. In previous measurements co-expansion with a heavier Ne gas was needed to obtain sufficiently low H$_2$ velocities and gain access to the regime in which resonances can be observed.\cite{Vogels:NatChem10:435} In the present work, this regime is reached by significantly reducing the beam intersection angle. We therefore can reach low collision energies at higher laboratory frame velocities, enabling the use of neat beams of H$_2$. Collisions involving both \textit{para}- and \textit{ortho}-H$_2$ were investigated separately, resolving resonance features for both systems. Measurements are compared with results calculated with the two previously mentioned PESs for the NO-H$_2$ system. Our results confirm the previous discernment between the two models, and show an excellent agreement with one of the two CCSD(T) potentials even in this low energy regime.

\section{Experimental Methods}

The experiments were conducted in a crossed-beam apparatus which has been described in detail previously.\cite{Jongh2020,Onvlee:PCCP16:15768} We briefly highlight its main aspects here. A gas consisting of 5\% NO seeded in a carrier gas - consisting of either Ar or an Ar/Ne mixture - was expanded through a Nijmegen Pulsed Valve at 1 bar backing pressure.\cite{Yan:RSI84:023102} After passing through a skimmer, the molecules entered a 316 stage Stark decelerator, operated in $s = 3$ mode at a guiding phase angle of $\phi_0 = 0\degree$, and at a voltage difference of 36 kV between opposite electrodes.\cite{Meerakker:CR112:4828} This allowed for state-selection of the NO molecules in their $X^2\Pi_{1/2}$, $v = 0$, $j = 1/2, f$ state [referred to hereafter as ($j = 1/2, f$)] at a controllable mean velocity ranging from 500 m/s to 880 m/s. Fine tuning of the Stark decelerator also resulted in a small temporal width of the NO packet, reaching a typical value of 15 $\mu$s as well as low velocity spreads of about 5 m/s. All spreads in this manuscript are given in full width at half maximum (FWHM) unless stated otherwise. At the exit of the Stark decelerator, the molecules traverse a distance of 529.5 mm in free flight and collide with the secondary H$_2$ beam under a 5.2$\degree$ beam intersection angle.
 
We generated the neat beam of H$_2$ molecules with a cryogenically cooled Even-Lavie valve\cite{Even:AiC2014:636042} by expanding the gas at backing pressures between 0.6 and 2 bar. The valve was operated at temperatures between 20 and 28 K, resulting in velocities ranging from 760 m/s to 840 m/s. The beam was collimated by a 50 mm long, 3 mm diameter skimmer placed at a distance of 162 mm from the nozzle of the valve. The distance between the nozzle of the Even-Lavie valve and the intersection region was 460 mm.

Due to the degeneracy of its nuclear spin levels, a normal beam of H$_2$ molecules consists of $25\%$ \textit{para}-H$_2$ and $75\%$ \textit{ortho}-H$_2$, in which the hydrogen molecules occupy the even and odd rotational states, respectively. Since conversion between \textit{ortho-} and \textit{para}-sublevels is spin-forbidden in the absence of a magnetic field, the molecular beam expansion ensures that only the lowest rotational levels, $j = 0$ (\textit{para}-H$_2$) and $1$ (\textit{ortho}-H$_2$), are occupied. Scattering experiments were performed with beams of pure \textit{para}-H$_2$ as well as with beams of normal H$_2$-gas. Pure \textit{para}-H$_2$ was made by liquefying the gas before expansion in the presence of Nickel(II)-Sulfate. This acts as a magnetic catalyst, resulting in efficient \textit{ortho-para}-conversion, converting the molecules to their true lowest rotational state $j = 0$. This beam allowed for direct probing of scattering between NO and H$_2$ $(j=0)$.  Results for \textit{ortho}-H$_2$ $(j=1)$ were obtained by subtracting signal acquired with the pure \textit{para}-H$_2$ beam from that acquired with the normal H$_2$ beam. The rotational state distributions of the hydrogen beams were probed by rotational spectroscopy with a (2+1) REMPI scheme using 201 nm photons. Indeed, only the $j=0$ level was occupied for the \textit{para}-H$_2$ beam, whereas the $j=0$ and $j=1$ levels were populated with a 1:3 ratio for the normal H$_2$ beam. The \textit{para}-H$_2$ beam was found to contain less than 2\% \textit{ortho}-H$_2$.

We state-selectively detected the scattered NO radicals by applying a (1 + 1’) REMPI scheme using two pulsed dye laser systems pumped by a single Nd:YAG laser. A 226 nm frequency-tripled dye laser excited the NO molecules via the (0,0) band of the $A^2\Sigma^+ \leftarrow X ^2\Pi$ transition. A frequency-doubled dye laser subsequently ionized the NO just above the threshold energy (328 nm). The time delay between the lasers was set to approximately 5 ns. Full Doppler range coverage of the scattered NO molecules was ensured by the small range of available post-collisional lab-frame velocities at low collision energies. After ionization, we used VMI to detect the NO ions.\cite{Eppink:RSI68:3477,Townsend:RSI74:2530} The ion optics consisted of a repeller plate combined with 15 cylindrical extractors. These accelerated the ions through a 1 m long grounded time-of-flight tube towards a microchannel plate (MCP) coupled to a phosphor screen. Resulting emission by the phosphor screen was recorded by a CCD camera. Acquiring this image over many experimental cycles leads to an image that reflects the DCS of the probed scattering process. With the current detector, applying event counting and centroiding algorithms, we were able to achieve a resolution of 0.58 m/s/pixel. ICSs were probed using the same detection system, but by acquiring the total signal independent of position. These measurements were typically performed with the VMI lenses out of focus to prevent detector saturation.

For DCS measurements we used laser powers of 30 $\mu$J and 1.0 mJ for the 226 nm and 328 nm beams, respectively. The MCP was mass-gated such that only NO radicals were detected. We ran the experiment at a 10 Hz repetition rate and a single scattering image was measured by accumulating signal between $10^5$ and $10^6$ laser shots. For the ICS measurements, laser powers of 200 $\mu$J and 2.0 mJ were used for the two lasers, respectively. The collision energy range was scanned in on average 0.2 cm$^{-1}$ intervals by tuning the NO packet velocity with the Stark decelerator using an automated cycle. Further details of the ICS measurements and the necessary signal corrections are described in Ref.\ \onlinecite{Jongh2020}.

We calibrated the secondary beam velocity by careful measurements of the beam intersection angle $\alpha$ and an ICS measurement of the threshold behavior for collisional excitation of NO to the $j = 3/2, e$ level which opens at a collision energy of 5.0 cm$^{-1}$. The latter allowed us to accurately determine the collision energy $E_\mathrm{coll}$ and collision energy spread $\Delta E_\mathrm{coll}$ by fitting the experimental data to the predicted threshold behavior of the channel -- convoluted with the experimental energy spread -- by using the velocity $v_2$ and velocity spread $\Delta v_2$ of the H$_2$ beam as fitting parameters:

\begin{equation}\label{eq:ecol}
E_\mathrm{coll} = \frac{1}{2}\mu(v_1^2 +v_2^2 -2v_1v_2\cos\alpha),
\end{equation}
and
\begin{eqnarray} \label{eq:decol}
\Delta E_\mathrm{coll}^2 =& \mu^2[(v_1-v_2\cos\alpha)^2\Delta{v_1}^2 + \\
\nonumber
&(v_2-v_1\cos\alpha)^2\Delta{v_2}^2 + (v_1v_2\sin\alpha)^2\Delta{\alpha}^2],
\end{eqnarray}
where $\mu$ stands for the reduced mass of the NO-H$_2$ system.  The velocity spread $\Delta v_1$ of NO is determined from VMI measurements on the incoming beam, and the spread in the collision angle $\Delta \alpha$ is determined from trajectory simulations.

\section{Theoretical Methods}
We performed coupled-channels calculations for the NO-H$_2$ system using a scattering program able to handle bimolecular open-shell systems, as previously described in Ref.\ \onlinecite{Jongh2017}. State-to-state integral and differential cross sections were computed for collision energies ranging from $0-10$ cm$^{-1}$ in steps of 0.01 cm$^{-1}$. Contributions of partial waves up to total angular momentum $\mathcal{J} = 121/2$ and rotational levels up to $j_\mathrm{NO} = 15/2$ and $j_\mathrm{H_2} = 3$ were included in the channel basis employed in the scattering calculations. We propagated the wave function on a radial grid ranging from $R = 4.5$ to $40$ a$_{0}$ with a grid spacing of 0.1 a$_0$. 

Calculations were performed with the two most sophisticated PESs available for this system. Both are computed with the CCSD(T) method. The first PES was developed by \citet{Klos2017} and is constructed using the explicitly correlated F12a method, scaling the contribution of triple excitations. The second, described in Ref.\ \cite{Jongh2017} used a complete basis set (CBS) extrapolation. We will refer to these PESs as the F12 and CBS potential, respectively.

\section{Results}

\begin{figure}
\includegraphics[width=\columnwidth]{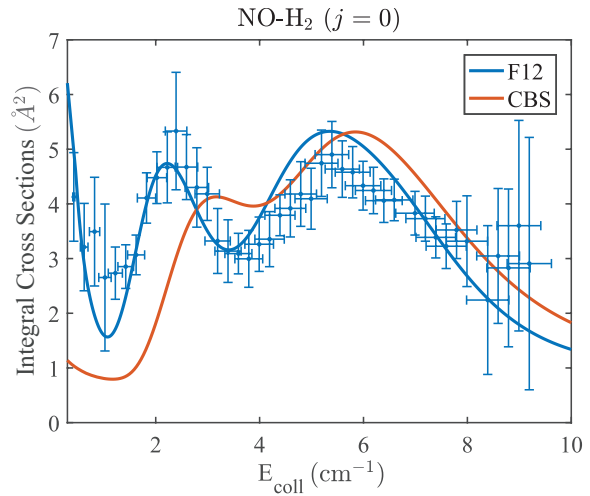}
\caption{Collision energy dependence of the ICS for the ($j = 1/2, f$) $\rightarrow$ ($j = 1/2, e$) channel of NO-H$_2$($j=0$) collisions. Measured cross sections (data points with error bars) are compared with those derived from the F12 (blue) and CBS (red) potentials. Experimental data are given in arbitrary units. Vertical error bars represent the uncertainties at the 95\% confidence level and horizontal error bars represent the standard errors in the mean collision energies, effectively displaying calibration uncertainties. The calculated cross sections were convoluted with experimental energy spreads, and the data points were vertically scaled to the theoretical curve of the F12 potential using root-mean-square fitting.}
\label{fig:ICS-j0}
\end{figure}

\begin{figure}
\includegraphics[width=\columnwidth]{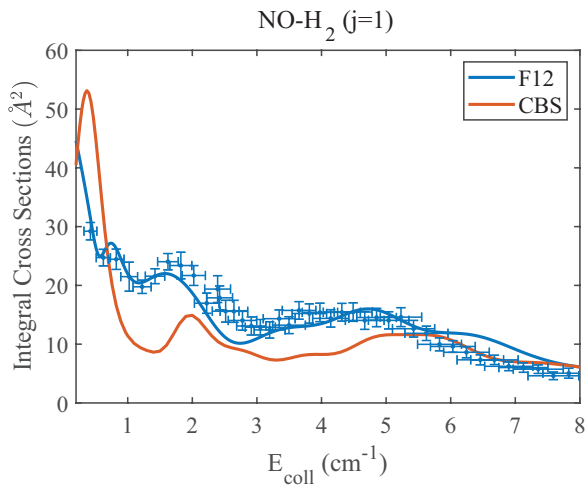}
\caption{Energy dependence of the ICS for the ($j = 1/2, f$) $\rightarrow$ ($j = 1/2, e$) channel of NO-H$_2$($j=1$) collisions. Theoretical ICSs resulting from both the F12 (blue) and CBS (red) potentials are shown. Experimental data points were vertically scaled to the theoretical curve of the F12 potential. See the caption of Fig.\ \ref{fig:ICS-j0} for more details.}
\label{fig:ICS-j1}
\end{figure}

We measured relative ICSs for the NO($j = 1/2, f$) + H$_{2}$ $\rightarrow$ NO($j = 1/2, e$) + H$_{2}$ parity changing process, which has an associated energy release of 0.01 cm$^{-1}$. Results for collisions involving \textit{para}-H$_2$ are plotted in Fig.\ \ref{fig:ICS-j0} and show two clear peaks associated with resonance behavior, as well as an incline at lower energies. We compared our experimental findings with ICSs computed from both the F12 and CBS potentials. To account for the experimental collision energy spread the theoretical cross sections were convoluted with a Gaussian distribution of variable width $\sigma(E)$ ranging from 0.03 to 1.6 cm$^{-1}$, depending on the collision energy. The resonance peaks from the F12 potential lie at lower energies than those predicted by the CBS potential. This originates from the fact that the F12 potential is deeper than the CBS potential by nearly 2.0 cm$^{-1}$. Results show an excellent agreement with predictions from the F12 potential.

Similar measurements were performed for collisions involving \textit{ortho}-H$_2$ as a scattering partner. As discussed, this data was extracted from measurements involving a 3:1 mixture of \textit{ortho}- and \textit{para}-H$_2$ by subtracting the measured ICS for NO-\textit{para}-H$_2$ in the appropriate ratio. Results are shown in Fig.\ \ref{fig:ICS-j1}. The anisotropic nature of the H$_2$ $(j = 1)$ results in completely different scattering behavior compared to the isotropic $(j = 0)$ case, but resonance features are again observed that better fit the predictions of the F12 potential.

\begin{figure}
\includegraphics[width=\columnwidth]{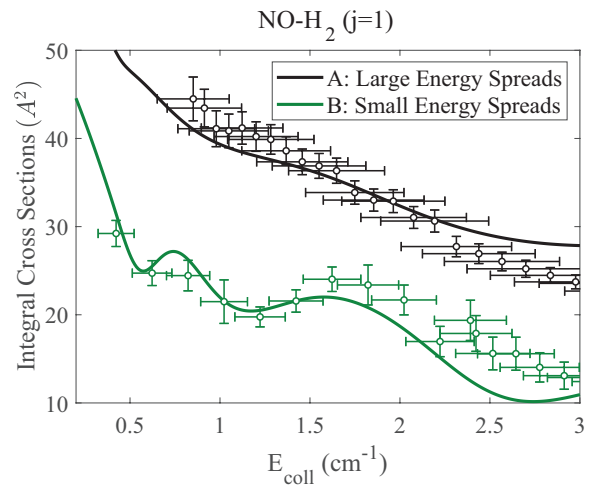}
\caption{Collision energy ($E_\mathrm{coll}$) dependence of ICS for the NO-H$_2$($j=1$) collision, measured with different energy spreads. Vertical error bars show the uncertainties at the 95\% confidence level. Horizontal error bars show the standard errors in the determination of the mean collision energies. The curves show the ICS calculated from the F12 potential and convoluted with the energy spreads of the corresponding measurements. Experimental data points were vertically scaled to the corresponding theoretical curves. ICS$_A$ has been given a vertical offset for clarity. See text for details.}
\label{fig:deltaE}
\end{figure}

\begin{table}
	\caption{Parameters for two sets of ICS measurements of the NO-H$_2$($j=1$) collision. See text for details.}
	\begin{tabularx}{\columnwidth}{YYY}
	\hline \hline
	\noalign{\vspace{0.1cm}}
	Parameter & ICS$_A$ & ICS$_B$ \\ 
	\hline 	
	\noalign{\vspace{0.1cm}}
	$v_1$ & 530--680 m/s & 600--790 m/s \\ [0.1cm]
	$\Delta{v_1}$ & 5 m/s & 5 m/s \\ [0.1cm]
	$T_2$ & 20 K & 23 K \\ [0.1cm]
	$p_2$ & 0.6 bar & 1.5 bar \\ [0.1cm]
	$v_2$ & 760 m/s & 790 m/s \\ [0.1cm]
	$\Delta{v_2}$ & 25 m/s & 15 m/s \\ [0.1cm]
	$\alpha$ & 5.2$\degree$ & 5.2$\degree$ \\ [0.1cm]
	$\Delta \alpha$ & 0.2$\degree$ & 0.2$\degree$ \\ [0.1cm]
	$E_\mathrm{coll}$ & 0.85--4.42 cm$^{-1}$ & 0.42--3.15 cm$^{-1}$ \\ [0.1cm]
	$\Delta E$ & 0.56--1.58 cm$^{-1}$ & 0.03--0.70 cm$^{-1}$ \\ 
	\hline \hline
	\end{tabularx}
\label{tab:deltaE}
\end{table}
Interestingly, theoretical predictions for collisions with \textit{ortho}-H$_2$ show a steep increase in the ICS at the lowest energies. The full range of this feature cannot be reached in the current experimental configuration as it requires access to collision energies below 0.4 cm$^{-1}$. Nevertheless, we were able to probe this region indirectly by tuning the experimental energy spreads to our advantage. From Eq. \ref{eq:ecol} it is clear that the same collision energy can be reached by different combinations of $v_1$ and $v_2$. According to Eq. \ref{eq:decol}, however, the involved spread differs for each combination. For most measurements we tune our parameters such that we obtain an optimized collision energy resolution -- as described in Ref.\ \onlinecite{Scharfenberg2011} -- but by deliberately increasing $\Delta E$ we can probe contributions from scattering events at energies a factor of two below our lowest achievable mean collision energy.

To achieve this we performed ICS measurements using two sets of $(v_1, v_2)$ combinations. Corresponding parameters are displayed in Table \ref{tab:deltaE}. The NO velocity was controlled by the Stark decelerator and the H$_2$ velocity was tuned by choosing an appropriate combination of backing pressure $p_2$ and Even-Lavie valve temperature $T_2$. We performed a measurement -- denoted as ICS$_\mathrm{A}$ -- with reduced energy resolution and compared our findings with our high-resolution measurements -- labeled as ICS$_\mathrm{B}$. Both measurements are shown in Fig.\ \ref{fig:deltaE} together with theoretical predictions based on the F12 potential, convoluted with the according energy spreads. 

We found that the two measurement series diverge at lower energies, with the decreased resolution measurement of ICS$_\mathrm{A}$ showing a stronger increase in scattering signal for decreasing energy. This indeed suggests the existence of a sharp resonance feature at energies below the experimental limit of 0.4 cm$^{-1}$. Both ICS$_\mathrm{A}$ and ICS$_\mathrm{B}$ are well captured by the theoretical curves, demonstrating not only the predictive ability of the F12 potential, but also the accuracy of our $E_\mathrm{coll}$ and $\Delta E_\mathrm{coll}$ calibrations.

\begin{figure}
	\includegraphics[width=\columnwidth]{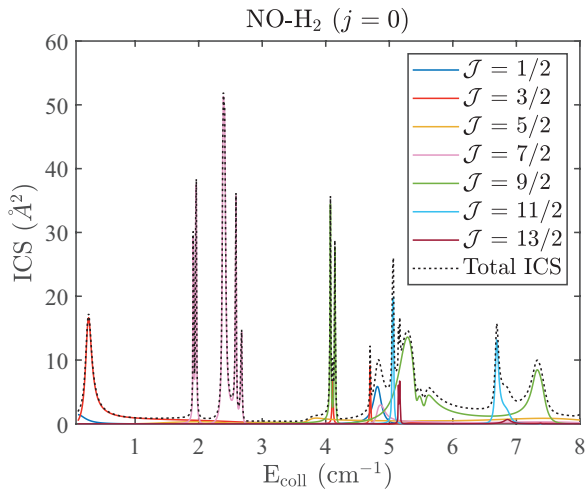}
	\caption{Theoretical partial cross sections based on the F12 potential for NO-H$_2$($j=0$) collisions in terms of total angular momentum $\mathcal{J}$.}
	\label{fig:Jcontrib-j0}
\end{figure}

\begin{figure}
	\includegraphics[width=\columnwidth]{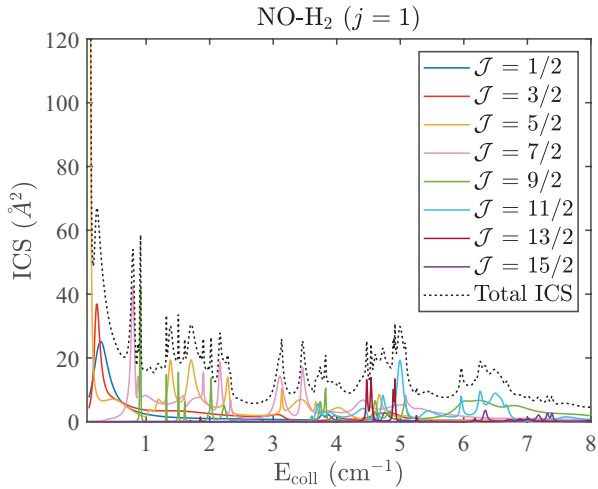}
	\caption{Theoretical partial cross sections based on the F12 potential for NO-H$_2$($j=1$) collisions in terms of total angular momentum $\mathcal{J}$. The sub-Kelvin regime is shown in more detail in Fig.\ \ref{fig:multipole}.}
	\label{fig:Jcontrib-j1}
\end{figure}

To further characterize the scattering behavior, we conducted a theoretical analysis on the partial wave nature of the observed resonances. By decomposing the cross sections in the contributions of individual partial waves we were able to assign a total angular momentum ($\mathcal{J}$) value -- a quantity conserved throughout the collision -- for all predicted  resonance features. Results based on the F12 potential for collisions of NO with \textit{para}- and \textit{ortho}-H$_2$ are shown in Fig.\ \ref{fig:Jcontrib-j0} and \ref{fig:Jcontrib-j1}, respectively.  For NO colliding with \textit{para}-H$_2$, at energies below 1 cm$^{-1}$ the interaction is dominated by a distinct resonance feature corresponding to $\mathcal{J} = 3/2$. At energies between 1 and 3 cm$^{-1}$, several sharp resonance features pertaining to $\mathcal{J} = 7/2$ can be clearly distinguished. At higher energies, several other resonance features related to different values of $\mathcal{J}$ can be seen. The partial wave decomposition for NO-\textit{ortho}-H$_2$ shows a strikingly different structure. Sub-Kelvin scattering is dominated by a vast resonance peak corresponding to  $\mathcal{J} = 5/2$, whereas a dense collection of resonance peaks is observed in the 1--8 cm$^{-1}$ region. The origin of the differences between scattering of NO with \textit{para}- or \textit{ortho}-H$_2$ will be discussed in section \ref{sec:discussion}.

\begin{figure}
\includegraphics[width=\columnwidth]{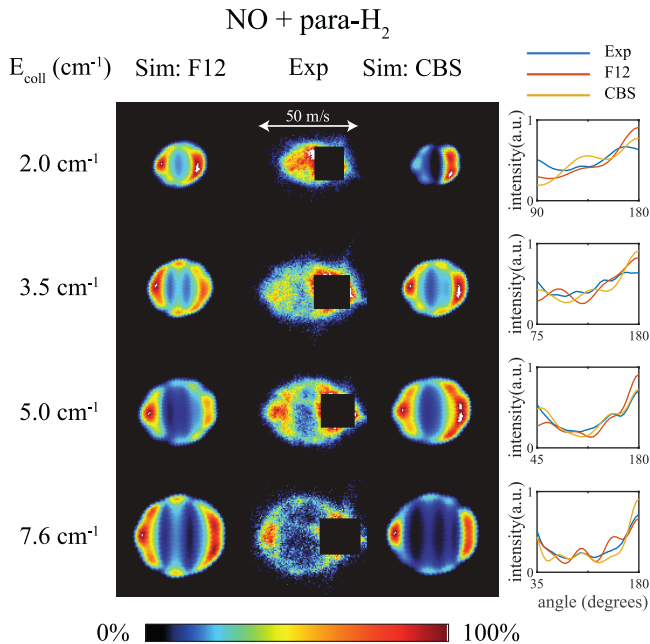}
\caption{Experimental and simulated ion images for NO-H$_2$($j=0$) collisions at several collision energies. Simulated ion images are shown in the left (based on the F12 potential) and right (based on the CBS potential) image columns. The center image column contains the experimental ion images. The relative velocity vector is horizontally oriented such that forward scattering signal is displayed on the right hand side of the image. A portion of the experimental image at forward scattering is masked due to imperfect state selection of the reagent NO packet. The angular distributions are derived from the experimental (blue) and simulated (red for F12, yellow for CBS potentials) images. They are displayed in the rightmost column for each respective collision energy. The curves are normalized with respect to their area.}
\label{fig:DCS-para}
\end{figure}

\begin{figure}
\includegraphics[width=\columnwidth]{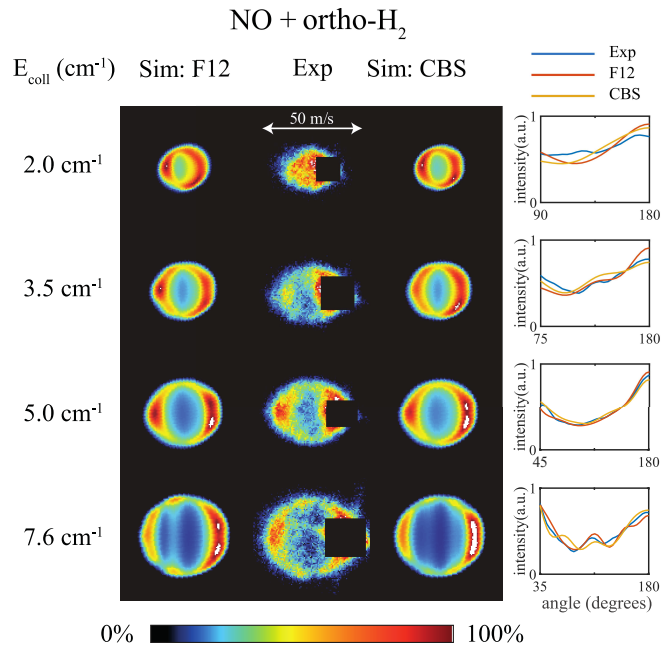}
\caption{Experimental and simulated ion images for NO-H$_2$($j=1$) collisions at several collision energies. See the caption of Fig.\ \ref{fig:DCS-para} for details.}
\label{fig:DCS-ortho}
\end{figure}

In order to elucidate the partial wave fingerprints of the observed resonances, we also measured the angular distributions of the scattered NO for both scattering systems at selected energies in the 2 -- 8 cm$^{-1}$ energy range. In previous work, we found that these angular distributions respond sensitively to the existence of resonances, and in some cases directly reflect the partial wave fingerprint underlying the resonances.\cite{Vogels:SCIENCE350:787,Vogels:NatChem10:435,Jongh2020} The experimental ion images for NO-H$_2(j=0)$ and NO-H$_2(j=1)$ are shown in Fig.\ \ref{fig:DCS-para} and \ref{fig:DCS-ortho}, respectively. Part of the forward scattering region is masked in the images due to imperfect state selection of NO. All images are normalized using the maximum intensity in the region used for the extraction of the angular distribution, with angular ranges depending on collision energy, as given in the respective figures.
At the lowest energies, the ion images appear slightly asymmetrical. This is attributed to the fact that reaching sufficient scattering signal for ion image measurements at these low energies requires a hydrogen beam with reduced speed ratio, as well as to aberrations in the VMI detector caused by the existence of stray magnetic fields.

Energy-dependent changes in the angular distributions are observed for both systems, which could be indicative of the presence of resonance features. All images furthermore display pronounced backward scattering. Simulated ion images were created with numerical trajectory simulations, as described previously \cite{Zastrow:NatChem6:216}, using DCSs provided by either the F12 or CBS potential as input. The results show good agreement with simulations for both potentials. Despite the vast differences in their respective ICSs, the angular distributions for NO colliding with either \textit{ortho}- or \textit{para}-H$_2$ show only subtle differences at the energies probed experimentally, and it is challenging to experimentally record scattering images with sufficient signal to noise ratio to probe these differences. Hence, for the systems and energies probed here, we find the unusual (and unexpected) situation that ICS measurements are more sensitive to resonance phenomena than DCS measurements. 

\section{Discussion}\label{sec:discussion}

Our measurements of ICS for scattering of NO with both \textit{para}-H$_2$ and \textit{ortho}-H$_2$ clearly show better agreement with results for the F12 potential compared to the CBS potential. We thus conclude that the better agreement for this potential found earlier\cite{Vogels:NatChem10:435} for energies down to 10 cm$^{-1}$  persists for energies down to 0.4 cm$^{-1}$. 

As discussed in detail in Ref.\ \onlinecite{Vogels:NatChem10:435}, the well in the F12 potential is about 2 cm$^{-1}$ deeper than the well in the CBS potential. As a result, low-energy resonances are predicted to occur at different energies by the two potentials; the differences are small yet significant enough to be distinguished in our ICS measurements.  Yet, this does not imply that the F12 potential is of higher quality than the CBS potential: as mentioned in Ref.\ \onlinecite{Vogels:NatChem10:435}, the deeper well for the F12 potential is most likely the result of the choice of basis set used in the calculations. Inclusion of a larger atomic basis would have caused its minimum to become less deep, so that it nearly coincides with the minimum in the CBS potential. Adding effects of quadruple excitations is expected to again lead to a deepening of the well, both for the F12 and CBS potentials. Such a CCSDT(Q) PES has recently been constructed for the NO-He system, and was found to be essential to yield satisfactory agreement with measurements of resonances in this system. \cite{Jongh2020} Unfortunately, due the high number of degrees of freedom for a bimolecular system, computation of such a potential for the NO-H$_2$ complex exceeds current limits. 

Still, the available PESs give us valuable insight into the nature of the low-energy scattering of this complex. Effects of partial wave resonances are observed, showing significant differences between collisions involving \textit{ortho}- and \textit{para}-H$_2$. These differences are expected to result from interactions arising from the coupling between hydrogen's quadrupole moment with the dipole and quadrupole moment of NO \cite{Klos2017}. They contribute only for H$_2$ in the $j=1$ state, since the $j=0$ state has a spherical charge distribution such that the interaction potential is governed exclusively by dispersion. The quadrupole-dipole and quadrupole-quadrupole interactions scale with $R^{-4}$ and $R^{-5}$, respectively, where $R$ is the distance between the center of masses of the two molecules. Since all other terms decay faster with $R$, these interactions dominate at long range and are therefore expected to govern low-energy scattering. Such effects were previously observed for Penning ionization reactions, showing stronger interactions for He$(2^3P_2)$ with \textit{ortho}-H$_2$ compared to \textit{para}-H$_2$.\cite{Shagam2015} For NO scattering with \textit{ortho}-H$_2$, the quadrupole moment of H$_2$ results in a much denser grid of bound-states compared to NO-\textit{para}-H$_2$ potential, explaining the denser cluster of resonances (quasi-bound states) observed for NO-\textit{ortho}-H$_2$ collisions \cite{Klos2017}. The contribution of the quadrupole-quadrupole interaction further explains the order of magnitude higher ICS oberved for \textit{ortho}- compared to \textit{para}-H$_2$ collisions at sub-Kelvin energies (see Fig.\ \ref{fig:Jcontrib-j0} and \ref{fig:Jcontrib-j1}). Additionally, most resonances for the NO-H$_{2}(j=1)$ system are found to be of Feshbach-type with the radial-wavefunctions being mixed with the higher lying spin-orbit states coupled by the off-diagonal quadrupole-quadrupole coupling term.

\begin{figure}
	\includegraphics[width=\columnwidth]{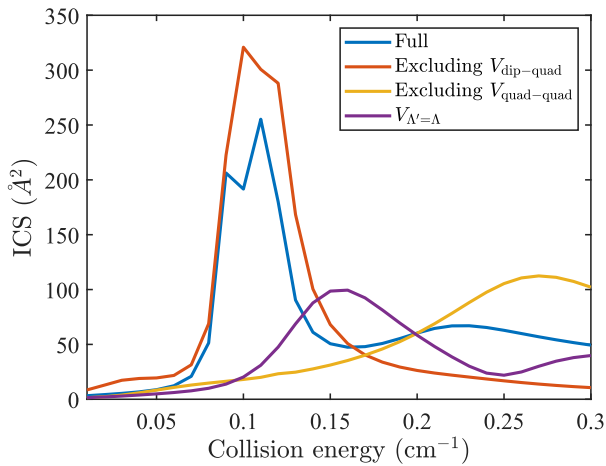}
	\caption{Theoretical cross section based on the F12 potential for the parity-changing channel of the NO-H$_2(j=1)$ interaction. Lines show contributions from different parts of the multipolar expansion to the ICS. The blue line corresponds to the full interaction, whereas for the red and yellow lines contributions from the dipole-quadrupole ($V_\mathrm{dip-quad}$) and quadrupole-quadrupole ($V_\mathrm{quad-quad}$) interactions are omitted, respectively. The purple line shows the ICS for diagonal ($\Lambda'=\Lambda$) interactions only.}
	\label{fig:multipole}
\end{figure}

To further investigate the contributions from the different potential terms to the ICS for NO-\textit{ortho}-H$_2$ collisions, we disentangled contributions from individual multipolar interactions using the multipolar expansion \cite{Wormer2005,Ma2014} and applied this to the F12 potential to compute the ICS for the sub-Kelvin resonance structure in NO-\textit{ortho}-H$_2$ while excluding either the dipole-quadrupole or quadrupole-quadrupole interactions. In addition, we computed a curve based only on diagonal contributions to the interaction potential  ($\Lambda'=\Lambda$, where $\Lambda^{(')}$ is the body-frame projection of the electronic orbital angular momentum of NO). The resulting ICSs are shown in Fig.\ \ref{fig:multipole} and clearly show that the resonance at 0.1 cm$^{-1}$ disappears when the quadrupole-quadrupole interaction is excluded, further illustrating that at these energies NO-H$_2(j=1)$ collisions are indeed dominated by interactions involving hydrogen's quadrupole moment.  

\section{Conclusions}
We presented a joint experimental and theoretical study of NO-H$_2$ collisions at energies down to 0.4 cm$^{-1}$. In a crossed beam setup, we measured the parity-changing de-excitation of NO for collisions with both \textit{para}-H$_2$ and \textit{ortho}-H$_2$, resolving resonance structures in the energy dependent cross sections. Furthermore, we observed the incline of a resonance structure at 0.1 cm$^{-1}$ in collisions between NO and \textit{ortho}-H$_2$ by advantageous tuning of the experimental energy resolution. We compared measurements of the ICS and DCS to theoretical predictions based on the two most advanced \textit{ab initio} PESs available for this system: the CBS and F12 potentials, which are both constructed using CCSD(T) methods. Previous work at higher energies showed a better fit of experimental data with the F12 potential\cite{Vogels:NatChem10:435}, which is deeper than the CBS potential by a mere 2 cm$^{-1}$, although from the computational methods there are reasons to assume that this agreement was rather fortuitous. Our present measurements shows that this agreement still holds for energies down to 0.4 cm$^{-1}$, suggesting that the true NO-H$_2$ potential closely resembles the current F12 potential. 

The scattering behavior at these low energies is dominated by long-range interactions involving the quadrupole moment of hydrogen. Effects of this quadrupole moment are averaged out in \textit{para}-H$_2 (j=0)$  collisions, which explains why the rotational ground state of hydrogen leads to an order of magnitude smaller ICS and vastly different resonance structures for sub-Kelvin collisions. Theoretically, we revealed the contributions to the collision of different multipole components of the NO-H$_2$ interaction potential, giving us relevant insight into the nature of sub-Kelvin bimolecular collisions.

\begin{acknowledgments}
This work is part of a research program of the Netherlands Organization for Scientific Research (NWO). The research leading to these results has received funding from the European Research Council (ERC) under the European Union's Seventh Framework Programme (FP7/2007-2013/ERC grant agreement nr. 335646 MOLBIL) and from the ERC under the European Union's Horizon 2020 Research and Innovation Program (grant agreement 817947 FICOMOL).
The expert technical support by Niek Janssen and Andr\'e van Roij is gratefully acknowledged. The authors thank Dr. Theo Cremers for developing data acquisition software.
\end{acknowledgments}

\section*{Data Availability}
The data that support the findings of this study are available from the corresponding author upon reasonable request.


\bibliographystyle{apsrev}
\bibliography{mp,string}

\begin{thebibliography}{34}
\expandafter\ifx\csname natexlab\endcsname\relax\def\natexlab#1{#1}\fi
\expandafter\ifx\csname bibnamefont\endcsname\relax
  \def\bibnamefont#1{#1}\fi
\expandafter\ifx\csname bibfnamefont\endcsname\relax
  \def\bibfnamefont#1{#1}\fi
\expandafter\ifx\csname citenamefont\endcsname\relax
  \def\citenamefont#1{#1}\fi
\expandafter\ifx\csname url\endcsname\relax
  \def\url#1{\texttt{#1}}\fi
\expandafter\ifx\csname urlprefix\endcsname\relax\def\urlprefix{URL }\fi
\providecommand{\bibinfo}[2]{#2}
\providecommand{\eprint}[2][]{\url{#2}}

\bibitem[{\citenamefont{Levine and Bernstein}(1987)}]{Levine:reaction-dynamics}
\bibinfo{author}{\bibfnamefont{R.~D.} \bibnamefont{Levine}} \bibnamefont{and}
  \bibinfo{author}{\bibfnamefont{R.~B.} \bibnamefont{Bernstein}},
  \emph{\bibinfo{title}{Molecular reaction dynamics and chemical reactivity}}
  (\bibinfo{publisher}{Oxford {U}niversity {P}ress, {N}ew {Y}ork},
  \bibinfo{year}{1987}).

\bibitem[{\citenamefont{Naulin and Costes}(2014)}]{Naulin:IRPC33:427}
\bibinfo{author}{\bibfnamefont{C.}~\bibnamefont{Naulin}} \bibnamefont{and}
  \bibinfo{author}{\bibfnamefont{M.}~\bibnamefont{Costes}},
  \bibinfo{journal}{Int. Rev. Phys. Chem.} \textbf{\bibinfo{volume}{33}},
  \bibinfo{pages}{427} (\bibinfo{year}{2014}).

\bibitem[{\citenamefont{Schutte et~al.}(1972)\citenamefont{Schutte, Bassi,
  Tommasini, and Scoles}}]{Schutte:PRL29:979}
\bibinfo{author}{\bibfnamefont{A.}~\bibnamefont{Schutte}},
  \bibinfo{author}{\bibfnamefont{D.}~\bibnamefont{Bassi}},
  \bibinfo{author}{\bibfnamefont{F.}~\bibnamefont{Tommasini}},
  \bibnamefont{and} \bibinfo{author}{\bibfnamefont{G.}~\bibnamefont{Scoles}},
  \bibinfo{journal}{Phys. Rev. Lett.} \textbf{\bibinfo{volume}{29}},
  \bibinfo{pages}{979} (\bibinfo{year}{1972}).

\bibitem[{\citenamefont{Grover et~al.}(1977)\citenamefont{Grover, Toennies,
  Welz, and Wolf}}]{Grover:ChemPhysLett:48:24}
\bibinfo{author}{\bibfnamefont{J.~R.} \bibnamefont{Grover}},
  \bibinfo{author}{\bibfnamefont{J.~P.} \bibnamefont{Toennies}},
  \bibinfo{author}{\bibfnamefont{W.}~\bibnamefont{Welz}}, \bibnamefont{and}
  \bibinfo{author}{\bibfnamefont{G.}~\bibnamefont{Wolf}},
  \bibinfo{journal}{Chem. Phys. Lett.} \textbf{\bibinfo{volume}{48}},
  \bibinfo{pages}{24} (\bibinfo{year}{1977}).

\bibitem[{\citenamefont{Toennies et~al.}(1979)\citenamefont{Toennies, Welz, and
  Wolf}}]{Toennies:JCP71:614}
\bibinfo{author}{\bibfnamefont{J.~P.} \bibnamefont{Toennies}},
  \bibinfo{author}{\bibfnamefont{W.}~\bibnamefont{Welz}}, \bibnamefont{and}
  \bibinfo{author}{\bibfnamefont{G.}~\bibnamefont{Wolf}}, \bibinfo{journal}{J.
  Chem. Phys.} \textbf{\bibinfo{volume}{71}}, \bibinfo{pages}{614}
  (\bibinfo{year}{1979}).

\bibitem[{\citenamefont{Chandler}(2010)}]{Chandler:JCP132:110901}
\bibinfo{author}{\bibfnamefont{D.~W.} \bibnamefont{Chandler}},
  \bibinfo{journal}{J. Comp. Phys.} \textbf{\bibinfo{volume}{132}},
  \bibinfo{pages}{110901} (\bibinfo{year}{2010}).

\bibitem[{\citenamefont{Skodje et~al.}(2000)\citenamefont{Skodje, Skouteris,
  Manolopoulos, Lee, Dong, and Liu}}]{Skodje:PRL85:1206}
\bibinfo{author}{\bibfnamefont{R.~T.} \bibnamefont{Skodje}},
  \bibinfo{author}{\bibfnamefont{D.}~\bibnamefont{Skouteris}},
  \bibinfo{author}{\bibfnamefont{D.~E.} \bibnamefont{Manolopoulos}},
  \bibinfo{author}{\bibfnamefont{S.-H.} \bibnamefont{Lee}},
  \bibinfo{author}{\bibfnamefont{F.}~\bibnamefont{Dong}}, \bibnamefont{and}
  \bibinfo{author}{\bibfnamefont{K.}~\bibnamefont{Liu}},
  \bibinfo{journal}{Phys. Rev. Lett.} \textbf{\bibinfo{volume}{85}},
  \bibinfo{pages}{1206} (\bibinfo{year}{2000}).

\bibitem[{\citenamefont{Shiu et~al.}(2004)\citenamefont{Shiu, Lin, and
  Liu}}]{Shiu:PRL92:103201}
\bibinfo{author}{\bibfnamefont{W.}~\bibnamefont{Shiu}},
  \bibinfo{author}{\bibfnamefont{J.~J.} \bibnamefont{Lin}}, \bibnamefont{and}
  \bibinfo{author}{\bibfnamefont{K.}~\bibnamefont{Liu}},
  \bibinfo{journal}{Phys. Rev. Lett.} \textbf{\bibinfo{volume}{92}},
  \bibinfo{eid}{103201} (\bibinfo{year}{2004}).

\bibitem[{\citenamefont{Qiu et~al.}(2006)\citenamefont{Qiu, Ren, Che, Dai,
  Harich, Wang, Yang, Xu, Xie, Gustafsson et~al.}}]{Qiu:Science311:1440}
\bibinfo{author}{\bibfnamefont{M.~H.} \bibnamefont{Qiu}},
  \bibinfo{author}{\bibfnamefont{Z.~F.} \bibnamefont{Ren}},
  \bibinfo{author}{\bibfnamefont{L.}~\bibnamefont{Che}},
  \bibinfo{author}{\bibfnamefont{D.~X.} \bibnamefont{Dai}},
  \bibinfo{author}{\bibfnamefont{S.~A.} \bibnamefont{Harich}},
  \bibinfo{author}{\bibfnamefont{X.~Y.} \bibnamefont{Wang}},
  \bibinfo{author}{\bibfnamefont{X.~M.} \bibnamefont{Yang}},
  \bibinfo{author}{\bibfnamefont{C.~X.} \bibnamefont{Xu}},
  \bibinfo{author}{\bibfnamefont{D.~Q.} \bibnamefont{Xie}},
  \bibinfo{author}{\bibfnamefont{M.}~\bibnamefont{Gustafsson}},
  \bibnamefont{et~al.}, \bibinfo{journal}{Science}
  \textbf{\bibinfo{volume}{311}}, \bibinfo{pages}{1440} (\bibinfo{year}{2006}).

\bibitem[{\citenamefont{Dong et~al.}(2010)\citenamefont{Dong, Xiao, Wang, Dai,
  Yang, and Zhang}}]{Dong:Science327:1501}
\bibinfo{author}{\bibfnamefont{W.}~\bibnamefont{Dong}},
  \bibinfo{author}{\bibfnamefont{C.}~\bibnamefont{Xiao}},
  \bibinfo{author}{\bibfnamefont{T.}~\bibnamefont{Wang}},
  \bibinfo{author}{\bibfnamefont{D.}~\bibnamefont{Dai}},
  \bibinfo{author}{\bibfnamefont{X.}~\bibnamefont{Yang}}, \bibnamefont{and}
  \bibinfo{author}{\bibfnamefont{D.~H.} \bibnamefont{Zhang}},
  \bibinfo{journal}{Science} \textbf{\bibinfo{volume}{327}},
  \bibinfo{pages}{1501} (\bibinfo{year}{2010}).

\bibitem[{\citenamefont{Wang et~al.}(2013)\citenamefont{Wang, Chen, Yang, Xiao,
  Sun, Huang, Dai, Yang, and Zhang}}]{Wang:Science342:1499}
\bibinfo{author}{\bibfnamefont{T.}~\bibnamefont{Wang}},
  \bibinfo{author}{\bibfnamefont{J.}~\bibnamefont{Chen}},
  \bibinfo{author}{\bibfnamefont{T.}~\bibnamefont{Yang}},
  \bibinfo{author}{\bibfnamefont{C.}~\bibnamefont{Xiao}},
  \bibinfo{author}{\bibfnamefont{Z.}~\bibnamefont{Sun}},
  \bibinfo{author}{\bibfnamefont{L.}~\bibnamefont{Huang}},
  \bibinfo{author}{\bibfnamefont{D.}~\bibnamefont{Dai}},
  \bibinfo{author}{\bibfnamefont{X.}~\bibnamefont{Yang}}, \bibnamefont{and}
  \bibinfo{author}{\bibfnamefont{D.~H.} \bibnamefont{Zhang}},
  \bibinfo{journal}{Science} \textbf{\bibinfo{volume}{342}},
  \bibinfo{pages}{1499} (\bibinfo{year}{2013}).

\bibitem[{\citenamefont{Henson et~al.}(2012)\citenamefont{Henson, Gersten,
  Shagam, Narevicius, and Narevicius}}]{Henson2012}
\bibinfo{author}{\bibfnamefont{A.~B.} \bibnamefont{Henson}},
  \bibinfo{author}{\bibfnamefont{S.}~\bibnamefont{Gersten}},
  \bibinfo{author}{\bibfnamefont{Y.}~\bibnamefont{Shagam}},
  \bibinfo{author}{\bibfnamefont{J.}~\bibnamefont{Narevicius}},
  \bibnamefont{and}
  \bibinfo{author}{\bibfnamefont{E.}~\bibnamefont{Narevicius}},
  \bibinfo{journal}{Science} \textbf{\bibinfo{volume}{338}},
  \bibinfo{pages}{234} (\bibinfo{year}{2012}).

\bibitem[{\citenamefont{Lavert-Ofir et~al.}(2014)\citenamefont{Lavert-Ofir,
  Shagam, Henson, Gersten, K{\l}os, {\.Z}uchowski, Narevicius, and
  Narevicius}}]{Lavert-Ofir:NatChem6:332}
\bibinfo{author}{\bibfnamefont{E.}~\bibnamefont{Lavert-Ofir}},
  \bibinfo{author}{\bibfnamefont{Y.}~\bibnamefont{Shagam}},
  \bibinfo{author}{\bibfnamefont{A.~B.} \bibnamefont{Henson}},
  \bibinfo{author}{\bibfnamefont{S.}~\bibnamefont{Gersten}},
  \bibinfo{author}{\bibfnamefont{J.}~\bibnamefont{K{\l}os}},
  \bibinfo{author}{\bibfnamefont{P.~S.} \bibnamefont{{\.Z}uchowski}},
  \bibinfo{author}{\bibfnamefont{J.}~\bibnamefont{Narevicius}},
  \bibnamefont{and}
  \bibinfo{author}{\bibfnamefont{E.}~\bibnamefont{Narevicius}},
  \bibinfo{journal}{Nat. Chem.} \textbf{\bibinfo{volume}{6}},
  \bibinfo{pages}{332} (\bibinfo{year}{2014}).

\bibitem[{\citenamefont{Jankunas et~al.}(2015)\citenamefont{Jankunas,
  Jachymski, Hapka, and Osterwalder}}]{Jankunas2015}
\bibinfo{author}{\bibfnamefont{J.}~\bibnamefont{Jankunas}},
  \bibinfo{author}{\bibfnamefont{K.}~\bibnamefont{Jachymski}},
  \bibinfo{author}{\bibfnamefont{M.}~\bibnamefont{Hapka}}, \bibnamefont{and}
  \bibinfo{author}{\bibfnamefont{A.}~\bibnamefont{Osterwalder}},
  \bibinfo{journal}{J. Chem. Phys.} \textbf{\bibinfo{volume}{142}},
  \bibinfo{pages}{164305} (\bibinfo{year}{2015}).

\bibitem[{\citenamefont{Shagam et~al.}(2015)\citenamefont{Shagam, Klein,
  Skomorowski, Yun, Averbukh, Koch, and Narevicius}}]{Shagam2015}
\bibinfo{author}{\bibfnamefont{Y.}~\bibnamefont{Shagam}},
  \bibinfo{author}{\bibfnamefont{A.}~\bibnamefont{Klein}},
  \bibinfo{author}{\bibfnamefont{W.}~\bibnamefont{Skomorowski}},
  \bibinfo{author}{\bibfnamefont{R.}~\bibnamefont{Yun}},
  \bibinfo{author}{\bibfnamefont{V.}~\bibnamefont{Averbukh}},
  \bibinfo{author}{\bibfnamefont{C.~P.} \bibnamefont{Koch}}, \bibnamefont{and}
  \bibinfo{author}{\bibfnamefont{E.}~\bibnamefont{Narevicius}},
  \bibinfo{journal}{Nat. Chem.} \textbf{\bibinfo{volume}{7}},
  \bibinfo{pages}{921} (\bibinfo{year}{2015}).

\bibitem[{\citenamefont{Klein et~al.}(2017)\citenamefont{Klein, Shagam,
  Skomorowski, {\.Z}uchowski, Pawlak, Janssen, Moiseyev, van~de Meerakker,
  van~der Avoird, Koch et~al.}}]{Klein:NatPhys13:35}
\bibinfo{author}{\bibfnamefont{A.}~\bibnamefont{Klein}},
  \bibinfo{author}{\bibfnamefont{Y.}~\bibnamefont{Shagam}},
  \bibinfo{author}{\bibfnamefont{W.}~\bibnamefont{Skomorowski}},
  \bibinfo{author}{\bibfnamefont{P.~S.} \bibnamefont{{\.Z}uchowski}},
  \bibinfo{author}{\bibfnamefont{M.}~\bibnamefont{Pawlak}},
  \bibinfo{author}{\bibfnamefont{L.~M.} \bibnamefont{Janssen}},
  \bibinfo{author}{\bibfnamefont{N.}~\bibnamefont{Moiseyev}},
  \bibinfo{author}{\bibfnamefont{S.~Y.} \bibnamefont{van~de Meerakker}},
  \bibinfo{author}{\bibfnamefont{A.}~\bibnamefont{van~der Avoird}},
  \bibinfo{author}{\bibfnamefont{C.~P.} \bibnamefont{Koch}},
  \bibnamefont{et~al.}, \bibinfo{journal}{Nat. Phys.}
  \textbf{\bibinfo{volume}{13}}, \bibinfo{pages}{35} (\bibinfo{year}{2017}).

\bibitem[{\citenamefont{Chefdeville et~al.}(2013)\citenamefont{Chefdeville,
  Kalugina, van~de Meerakker, Naulin, Lique, and
  Costes}}]{Chefdeville:Science341:06092013}
\bibinfo{author}{\bibfnamefont{S.}~\bibnamefont{Chefdeville}},
  \bibinfo{author}{\bibfnamefont{Y.}~\bibnamefont{Kalugina}},
  \bibinfo{author}{\bibfnamefont{S.~Y.~T.} \bibnamefont{van~de Meerakker}},
  \bibinfo{author}{\bibfnamefont{C.}~\bibnamefont{Naulin}},
  \bibinfo{author}{\bibfnamefont{F.}~\bibnamefont{Lique}}, \bibnamefont{and}
  \bibinfo{author}{\bibfnamefont{M.}~\bibnamefont{Costes}},
  \bibinfo{journal}{Science} \textbf{\bibinfo{volume}{341}},
  \bibinfo{pages}{1094} (\bibinfo{year}{2013}).

\bibitem[{\citenamefont{Bergeat et~al.}(2015)\citenamefont{Bergeat, Onvlee,
  Naulin, van~der Avoird, and Costes}}]{Bergeat:NatChem7:349}
\bibinfo{author}{\bibfnamefont{A.}~\bibnamefont{Bergeat}},
  \bibinfo{author}{\bibfnamefont{J.}~\bibnamefont{Onvlee}},
  \bibinfo{author}{\bibfnamefont{C.}~\bibnamefont{Naulin}},
  \bibinfo{author}{\bibfnamefont{A.}~\bibnamefont{van~der Avoird}},
  \bibnamefont{and} \bibinfo{author}{\bibfnamefont{M.}~\bibnamefont{Costes}},
  \bibinfo{journal}{Nat. Chem.} \textbf{\bibinfo{volume}{7}},
  \bibinfo{pages}{349} (\bibinfo{year}{2015}).

\bibitem[{\citenamefont{Chefdeville et~al.}(2015)\citenamefont{Chefdeville,
  Stoecklin, Naulin, Jankowski, Szalewicz, Faure, Costes, and
  Bergeat}}]{Chefdeville:AJL799:L9}
\bibinfo{author}{\bibfnamefont{S.}~\bibnamefont{Chefdeville}},
  \bibinfo{author}{\bibfnamefont{T.}~\bibnamefont{Stoecklin}},
  \bibinfo{author}{\bibfnamefont{C.}~\bibnamefont{Naulin}},
  \bibinfo{author}{\bibfnamefont{P.}~\bibnamefont{Jankowski}},
  \bibinfo{author}{\bibfnamefont{K.}~\bibnamefont{Szalewicz}},
  \bibinfo{author}{\bibfnamefont{A.}~\bibnamefont{Faure}},
  \bibinfo{author}{\bibfnamefont{M.}~\bibnamefont{Costes}}, \bibnamefont{and}
  \bibinfo{author}{\bibfnamefont{A.}~\bibnamefont{Bergeat}},
  \bibinfo{journal}{Astr. J. Lett.} \textbf{\bibinfo{volume}{799}},
  \bibinfo{pages}{L9} (\bibinfo{year}{2015}).

\bibitem[{\citenamefont{Vogels et~al.}(2015)\citenamefont{Vogels, Onvlee,
  Chefdeville, van~der Avoird, Groenenboom, and van~de
  Meerakker}}]{Vogels:SCIENCE350:787}
\bibinfo{author}{\bibfnamefont{S.~N.} \bibnamefont{Vogels}},
  \bibinfo{author}{\bibfnamefont{J.}~\bibnamefont{Onvlee}},
  \bibinfo{author}{\bibfnamefont{S.}~\bibnamefont{Chefdeville}},
  \bibinfo{author}{\bibfnamefont{A.}~\bibnamefont{van~der Avoird}},
  \bibinfo{author}{\bibfnamefont{G.~C.} \bibnamefont{Groenenboom}},
  \bibnamefont{and} \bibinfo{author}{\bibfnamefont{S.~Y.~T.}
  \bibnamefont{van~de Meerakker}}, \bibinfo{journal}{Science}
  \textbf{\bibinfo{volume}{350}}, \bibinfo{pages}{787} (\bibinfo{year}{2015}).

\bibitem[{\citenamefont{Vogels et~al.}(2018)\citenamefont{Vogels, Karman,
  K{\l}os, Besemer, Onvlee, van~der Avoird, Groenenboom, and van~de
  Meerakker}}]{Vogels:NatChem10:435}
\bibinfo{author}{\bibfnamefont{S.~N.} \bibnamefont{Vogels}},
  \bibinfo{author}{\bibfnamefont{T.}~\bibnamefont{Karman}},
  \bibinfo{author}{\bibfnamefont{J.}~\bibnamefont{K{\l}os}},
  \bibinfo{author}{\bibfnamefont{M.}~\bibnamefont{Besemer}},
  \bibinfo{author}{\bibfnamefont{J.}~\bibnamefont{Onvlee}},
  \bibinfo{author}{\bibfnamefont{A.}~\bibnamefont{van~der Avoird}},
  \bibinfo{author}{\bibfnamefont{G.~C.} \bibnamefont{Groenenboom}},
  \bibnamefont{and} \bibinfo{author}{\bibfnamefont{S.~Y.~T.}
  \bibnamefont{van~de Meerakker}}, \bibinfo{journal}{Nat. Chem.}
  \textbf{\bibinfo{volume}{10}}, \bibinfo{pages}{435} (\bibinfo{year}{2018}).

\bibitem[{\citenamefont{de~Jongh et~al.}(2020)\citenamefont{de~Jongh, Besemer,
  Shuai, Karman, van~der Avoird, Groenenboom, and van~de
  Meerakker}}]{Jongh2020}
\bibinfo{author}{\bibfnamefont{T.}~\bibnamefont{de~Jongh}},
  \bibinfo{author}{\bibfnamefont{M.}~\bibnamefont{Besemer}},
  \bibinfo{author}{\bibfnamefont{Q.}~\bibnamefont{Shuai}},
  \bibinfo{author}{\bibfnamefont{T.}~\bibnamefont{Karman}},
  \bibinfo{author}{\bibfnamefont{A.}~\bibnamefont{van~der Avoird}},
  \bibinfo{author}{\bibfnamefont{G.~C.} \bibnamefont{Groenenboom}},
  \bibnamefont{and} \bibinfo{author}{\bibfnamefont{S.~Y.~T.}
  \bibnamefont{van~de Meerakker}}, \bibinfo{journal}{Science}
  \textbf{\bibinfo{volume}{368}}, \bibinfo{pages}{626} (\bibinfo{year}{2020}).

\bibitem[{\citenamefont{Onvlee et~al.}(2014)\citenamefont{Onvlee, Vogels, von
  Zastrow, Parker, and van~de Meerakker}}]{Onvlee:PCCP16:15768}
\bibinfo{author}{\bibfnamefont{J.}~\bibnamefont{Onvlee}},
  \bibinfo{author}{\bibfnamefont{S.~N.} \bibnamefont{Vogels}},
  \bibinfo{author}{\bibfnamefont{A.}~\bibnamefont{von Zastrow}},
  \bibinfo{author}{\bibfnamefont{D.~H.} \bibnamefont{Parker}},
  \bibnamefont{and} \bibinfo{author}{\bibfnamefont{S.~Y.~T.}
  \bibnamefont{van~de Meerakker}}, \bibinfo{journal}{Phys. Chem. Chem. Phys.}
  \textbf{\bibinfo{volume}{16}}, \bibinfo{pages}{15768} (\bibinfo{year}{2014}).

\bibitem[{\citenamefont{Yan et~al.}(2013)\citenamefont{Yan, Claus, van
  Oorschot, Gerritsen, Eppink, van~de Meerakker, and
  Parker}}]{Yan:RSI84:023102}
\bibinfo{author}{\bibfnamefont{B.}~\bibnamefont{Yan}},
  \bibinfo{author}{\bibfnamefont{P.~F.~H.} \bibnamefont{Claus}},
  \bibinfo{author}{\bibfnamefont{B.~G.~M.} \bibnamefont{van Oorschot}},
  \bibinfo{author}{\bibfnamefont{L.}~\bibnamefont{Gerritsen}},
  \bibinfo{author}{\bibfnamefont{A.~T. J.~B.} \bibnamefont{Eppink}},
  \bibinfo{author}{\bibfnamefont{S.~Y.~T.} \bibnamefont{van~de Meerakker}},
  \bibnamefont{and} \bibinfo{author}{\bibfnamefont{D.~H.}
  \bibnamefont{Parker}}, \bibinfo{journal}{Rev. Sci. Inst.}
  \textbf{\bibinfo{volume}{84}}, \bibinfo{eid}{023102} (\bibinfo{year}{2013}).

\bibitem[{\citenamefont{van~de Meerakker et~al.}(2012)\citenamefont{van~de
  Meerakker, Bethlem, Vanhaecke, and Meijer}}]{Meerakker:CR112:4828}
\bibinfo{author}{\bibfnamefont{S.~Y.~T.} \bibnamefont{van~de Meerakker}},
  \bibinfo{author}{\bibfnamefont{H.~L.} \bibnamefont{Bethlem}},
  \bibinfo{author}{\bibfnamefont{N.}~\bibnamefont{Vanhaecke}},
  \bibnamefont{and} \bibinfo{author}{\bibfnamefont{G.}~\bibnamefont{Meijer}},
  \bibinfo{journal}{Chem. Rev.} \textbf{\bibinfo{volume}{112}},
  \bibinfo{pages}{4828} (\bibinfo{year}{2012}).

\bibitem[{\citenamefont{Even}(2014)}]{Even:AiC2014:636042}
\bibinfo{author}{\bibfnamefont{U.}~\bibnamefont{Even}},
  \bibinfo{journal}{Advances in Chemistry} \textbf{\bibinfo{volume}{2014}},
  \bibinfo{pages}{636042} (\bibinfo{year}{2014}).

\bibitem[{\citenamefont{Eppink and Parker}(1997)}]{Eppink:RSI68:3477}
\bibinfo{author}{\bibfnamefont{A.~T. J.~B.} \bibnamefont{Eppink}}
  \bibnamefont{and} \bibinfo{author}{\bibfnamefont{D.~H.}
  \bibnamefont{Parker}}, \bibinfo{journal}{Rev. Sci. Inst.}
  \textbf{\bibinfo{volume}{68}}, \bibinfo{pages}{3477} (\bibinfo{year}{1997}).

\bibitem[{\citenamefont{Townsend et~al.}(2003)\citenamefont{Townsend, Minitti,
  and Suits}}]{Townsend:RSI74:2530}
\bibinfo{author}{\bibfnamefont{D.}~\bibnamefont{Townsend}},
  \bibinfo{author}{\bibfnamefont{M.~P.} \bibnamefont{Minitti}},
  \bibnamefont{and} \bibinfo{author}{\bibfnamefont{A.~G.} \bibnamefont{Suits}},
  \bibinfo{journal}{Rev. Sci. Inst.} \textbf{\bibinfo{volume}{74}},
  \bibinfo{pages}{2530} (\bibinfo{year}{2003}).

\bibitem[{\citenamefont{de~Jongh et~al.}(2017)\citenamefont{de~Jongh, Karman,
  Vogels, Besemer, Onvlee, Suits, Thompson, Groenenboom, van~der Avoird, and
  van~de Meerakker}}]{Jongh2017}
\bibinfo{author}{\bibfnamefont{T.}~\bibnamefont{de~Jongh}},
  \bibinfo{author}{\bibfnamefont{T.}~\bibnamefont{Karman}},
  \bibinfo{author}{\bibfnamefont{S.~N.} \bibnamefont{Vogels}},
  \bibinfo{author}{\bibfnamefont{M.}~\bibnamefont{Besemer}},
  \bibinfo{author}{\bibfnamefont{J.}~\bibnamefont{Onvlee}},
  \bibinfo{author}{\bibfnamefont{A.~G.} \bibnamefont{Suits}},
  \bibinfo{author}{\bibfnamefont{J.~O.~F.} \bibnamefont{Thompson}},
  \bibinfo{author}{\bibfnamefont{G.~C.} \bibnamefont{Groenenboom}},
  \bibinfo{author}{\bibfnamefont{A.}~\bibnamefont{van~der Avoird}},
  \bibnamefont{and} \bibinfo{author}{\bibfnamefont{S.~Y.~T.}
  \bibnamefont{van~de Meerakker}}, \bibinfo{journal}{J. Chem. Phys.}
  \textbf{\bibinfo{volume}{147}}, \bibinfo{pages}{013918}
  (\bibinfo{year}{2017}).

\bibitem[{\citenamefont{K{\l}os et~al.}(2017)\citenamefont{K{\l}os, Ma,
  Alexander, and Dagdigian}}]{Klos2017}
\bibinfo{author}{\bibfnamefont{J.}~\bibnamefont{K{\l}os}},
  \bibinfo{author}{\bibfnamefont{Q.}~\bibnamefont{Ma}},
  \bibinfo{author}{\bibfnamefont{M.~H.} \bibnamefont{Alexander}},
  \bibnamefont{and} \bibinfo{author}{\bibfnamefont{P.~J.}
  \bibnamefont{Dagdigian}}, \bibinfo{journal}{J. Chem. Phys}
  \textbf{\bibinfo{volume}{146}}, \bibinfo{pages}{114301}
  (\bibinfo{year}{2017}).

\bibitem[{\citenamefont{Scharfenberg et~al.}(2011)\citenamefont{Scharfenberg,
  van~de Meerakker, and Meijer}}]{Scharfenberg2011}
\bibinfo{author}{\bibfnamefont{L.}~\bibnamefont{Scharfenberg}},
  \bibinfo{author}{\bibfnamefont{S.~Y.~T.} \bibnamefont{van~de Meerakker}},
  \bibnamefont{and} \bibinfo{author}{\bibfnamefont{G.}~\bibnamefont{Meijer}},
  \bibinfo{journal}{Phys. Chem. Chem. Phys.} \textbf{\bibinfo{volume}{13}},
  \bibinfo{pages}{8448} (\bibinfo{year}{2011}).

\bibitem[{\citenamefont{von Zastrow et~al.}(2014)\citenamefont{von Zastrow,
  Onvlee, Vogels, Groenenboom, van~der Avoird, and van~de
  Meerakker}}]{Zastrow:NatChem6:216}
\bibinfo{author}{\bibfnamefont{A.}~\bibnamefont{von Zastrow}},
  \bibinfo{author}{\bibfnamefont{J.}~\bibnamefont{Onvlee}},
  \bibinfo{author}{\bibfnamefont{S.~N.} \bibnamefont{Vogels}},
  \bibinfo{author}{\bibfnamefont{G.~C.} \bibnamefont{Groenenboom}},
  \bibinfo{author}{\bibfnamefont{A.}~\bibnamefont{van~der Avoird}},
  \bibnamefont{and} \bibinfo{author}{\bibfnamefont{S.~Y.~T.}
  \bibnamefont{van~de Meerakker}}, \bibinfo{journal}{Nat. Chem.}
  \textbf{\bibinfo{volume}{6}}, \bibinfo{pages}{216} (\bibinfo{year}{2014}).

\bibitem[{\citenamefont{Wormer et~al.}(2005)\citenamefont{Wormer, K{\l}os,
  Groenenboom, and van~der Avoird}}]{Wormer2005}
\bibinfo{author}{\bibfnamefont{P.~E.} \bibnamefont{Wormer}},
  \bibinfo{author}{\bibfnamefont{J.~A.} \bibnamefont{K{\l}os}},
  \bibinfo{author}{\bibfnamefont{G.~C.} \bibnamefont{Groenenboom}},
  \bibnamefont{and} \bibinfo{author}{\bibfnamefont{A.}~\bibnamefont{van~der
  Avoird}}, \bibinfo{journal}{J. Chem. Phys.} \textbf{\bibinfo{volume}{122}},
  \bibinfo{pages}{244325} (\bibinfo{year}{2005}).

\bibitem[{\citenamefont{Ma et~al.}(2014)\citenamefont{Ma, K{\l}os, Alexander,
  van~der Avoird, and Dagdigian}}]{Ma2014}
\bibinfo{author}{\bibfnamefont{Q.}~\bibnamefont{Ma}},
  \bibinfo{author}{\bibfnamefont{J.}~\bibnamefont{K{\l}os}},
  \bibinfo{author}{\bibfnamefont{M.~H.} \bibnamefont{Alexander}},
  \bibinfo{author}{\bibfnamefont{A.}~\bibnamefont{van~der Avoird}},
  \bibnamefont{and} \bibinfo{author}{\bibfnamefont{P.~J.}
  \bibnamefont{Dagdigian}}, \bibinfo{journal}{J. Chem. Phys.}
  \textbf{\bibinfo{volume}{141}}, \bibinfo{pages}{174309}
  (\bibinfo{year}{2014}).

\end{thebibliography}

\end{document}